# The impact of reference isocentre position on set-up errors in head-and-neck image guided radiotherapy

Helena Lenko, BSc [1]

Primož Peterlin, PhD [1,*]

[1]Institute of oncology Ljubljana, Zaloška c. 2, SI-1000 Ljubljana, Slovenia

*Corresponding author; phone: +386-1-5879509, email: ppeterlin@onko-i.si

PP acknowledges the financial support from the Slovenian Research Agency (research core funding No. P1-0389).

# Abstract

**Aim**: To examine and quantify set-up errors in patient positioning in head-and-neck radiotherapy and to investigate the impact of the choice of reference isocentre – on the patient neck or patient skull – on the magnitude of set-up errors.

**Materials and methods**: Set-up position corrections obtained using online kV 2D/2D matching were recorded automatically for every treatment fraction. 3413 treatment records for 117 patients treated with volumetric modulated arc therapy (VMAT) during 2013 and 2014 on a single treatment machine in our clinic were analysed. In 79 treatment plans the reference isocentre was set to the patient skull, and in 47 to the neck.

**Results**: Standard deviation of group systematic error in the vertical, longitudinal and lateral direction and the couch rotation were found to be 2.5 mm, 2.1 mm, 1.9 mm and 0.43° (skull) and 2.5 mm, 1.8 mm, 1.7 mm and 0.49° (neck). Random error of the vertical, longitudinal, lateral and rotational position correction was 1.8 mm, 1.5 mm, 1.6 mm and 0.62° (skull) and 1.9 mm, 1.6 mm, 1.5 mm and 0.60° (neck). Positional shifts in different directions were found to be uncorrelated.

**Conclusions**: Neither reference isocentre set-up shows a clear advantage over the other in terms of interfraction set-up error.

# Background and aim

One of the key steps in the computed tomography (CT) simulation process is setting the reference isocentre (sometimes also called the patient origin)(1). Reference isocentre is a point fixed to the patient body, relative to which the treatment field isocentre (when isocentric technique is used) is defined. The reference isocentre is generally neither the Digital Imaging and Communications in Medicine (DICOM) origin, which is defined with respect to the CT scanner, nor the treatment field isocentre, which is not yet known at the time of CT simulation. A well-chosen reference isocentre however fulfills several criteria: (a) it is placed over stable anatomical structures, and (b) it lies as close as possible to the presumed treatment field isocentre.

For head-and-neck radiotherapy, reference isocentre is set either to the patient skull, with the radio-opaque markers positioned over the frontal and both parietal bones, or to the patient neck, usually on a transversal plane aligned with the cranial edge of the thyroid cartilage. The former has the advantage of being placed over a very stable anatomical structure, but it lies further away from the treatment field isocentre. Placing the reference isocentre onto the patient neck has the advantage of being closer to the treatment field isocentre, but this point may not be accessible, e.g., when tracheotomy was performed. The purpose of this study is to determine whether either choice of reference isocentre is advantageous to the other with respect to patient set-up errors, and consequently establish the priority of reference positioning criteria.

# Materials and metods

## Patient selection

Set-up correction data for 117 patients with head-and-neck cancer treated with volumetric modulated arc therapy (VMAT) in the years 2013 and 2014 on a single treatment machine (Varian NovalisTx; Varian Medical Systems, Palo Alto, CA, USA) in our clinic were analysed. This cohort represents all head-and-neck patients treated on this treatment machine during this period. 113 patients were treated with a radical intent.

## Patient set-up

All patients were immobilised with a thermoplastic mask (Posicast 5-point head, neck and shoulder mask; Civco Medical Solutions, Orange City, IA, USA), radio-opaque markers were taped to the thermoplastic shell to mark the anterior and both lateral laser cross-hairs designating the reference isocentre, and a planning CT scan was taken on a CT scanner with a flat-top couch. Radiation oncologists prescribed the dose and did the contouring of the targets and the organs-at-risk, upon which a VMAT treatment plan was created by the dosimetrists and/or medical physicists. Eclipse v10.0 treatment planning system (Varian Medical Systems, Palo Alto, CA, USA) was used.

An on-line imaging protocol with daily imaging was employed for the treatment. At every treatment fraction, a patient was positioned on the treatment couch using the markers as guidance so that the patient reference isocentre was aligned with the in-room lasers. Then, the planned shifts in the longitudinal, lateral and vertical direction were performed, bringing the patient isocentre in alignment with the in-room lasers. Two orthogonal kV images were taken using the on-board kV imager (On-Board Imager (OBI); Varian Medical Systems, Palo Alto, CA, USA). Focusing on bone anatomy landmarks, the radiation therapist in charge compared the alignment with the digitally reconstructed radiographs (DRR), determined the necessary corrections in patient position and refined patient position. The patients were randomly assigned to the radiation therapists on duty.

## Data retrieval

All the data concerning the positional corrections applied at every treatment fraction are recorded in the ARIA Oncology Information System (Varian Medical Systems). Using the Statistics/Trends tool in the Offline Review application, the positional corrections (Online OBI Match Results) were exported to a text file. Further processing was performed using in-house scripts written in GNU R (2).

## Systematic and random error

In order to reduce the set-up error, it is necessary to determine the systematic and the random component (3). Each of the $P$ patients in the study receives $n_k$ treatment fractions, with $k = 1, 2, ... P$. Let $x$ denote the positional correction – either couch shift (longitudinal, lateral, or vertical) or couch rotation (yaw). In the notation used, the first index of $x$ denotes the patient and the second the treatment fraction.

For each patient, the average positional shift and its standard deviation are calculated first:

$$\bar{x}_k = \frac{1}{n_k} \sum_{j=1}^{n_k} x_{kj} \qquad (1)$$

$$\sigma_k = \sqrt{\frac{1}{n_k - 1} \sum_{j=1}^{n_k} \left( x_{kj} - \bar{x}_k \right)^2} \qquad (2)$$

From average positional shifts, systematic error is calculated, and from standard deviations, the random error (4):

1. *Group systematic error*, or overall mean (denoted by $M$) is the average of individual averages $x_k$ calculated in Eq. 1. As the number of treatment fractions can differ from patient to patient, a weighted average must be used, the weight being the ratio between the number of treatment fractions for a given patient ($n_k$) and the average value of treatment fractions $n$. The latter is calculated as the ratio between the total number of fractions in all patients,

   $N = \sum_{k=1}^{P} n_k$ and the number of patients $P$, $n = N/P$:

   $$M = \frac{1}{P} \sum_{k=1}^{P} \frac{n_k}{\bar{n}} x_k = \frac{1}{N} \sum_{k=1}^{P} n_k \bar{x}_k \qquad (3)$$

2. *Standard deviation of group systematic error* is an estimate of the spread of individual average positional shifts $x_k$ around the overall mean $M$; the expression is similar to Eq. 2, only with (*P*-1) as normalization factor; again, a different number of fractions is accounted for with a weighting factor $n_k/n$:

$$\Sigma = \sqrt{\frac{1}{P-1}\sum_{k=1}^{P}\frac{n_k}{n}(\bar{x}_k - M)^2} = \sqrt{\frac{P}{N(P-1)}\sum_{k=1}^{P}n_k(\bar{x}_k - M)^2} \qquad (4)$$

Because of the finite sample size used in computing $x_k$, $\Sigma$ contains a random component. Correcting for this effect, one can define a corrected group systematic error $\Sigma'$:

$$\Sigma' = \sqrt{\Sigma^2 - \frac{\sigma^2}{\bar{n}}} \qquad (5)$$

3. *Random error*, or the population average of the standard deviation of random errors, is an estimate of the overall standard deviation; it is given by the root mean square of individual random errors $\sigma_k$ weighted with their respective number of degrees of freedom ($n_k - 1$):

$$\sigma = \sqrt{\frac{1}{N-P}\sum_{k=1}^{P}(n_k - 1)\sigma_k^2} \qquad (6)$$

The relation $\sum_{k=1}^{P}(n_k - 1) = N - P$ was used.

4. *Total error* is a measure of the total variation around the overall mean,

$$s = \sqrt{\frac{1}{N-1}\sum_{k=1}^{P}\sum_{j=1}^{n_k}(x_{kj} - M)^2} \qquad (7)$$

Here, index *k* runs over patients, and index *j* over treatment fractions. Total error can be calculated as the random error and the corrected systematic error combined, $s = \sqrt{\sigma^2 + \Sigma'^2}$.

# Results

Altogether, 3413 records on positional corrections for 126 treatment plans created for 117 patients were analysed. The number of treatment plans exceeds the number of patients because 9 patients needed re-planning during the course of the treatment. 952 records contained gross errors – positional displacement larger than 5 mm. 10 records contained positional displacement larger than 15 mm and were excluded from the further analysis.

In 79 of the treatment plans, reference isocentre was set on the patient skull, and in 47 of them, on the patient neck. Fig. 1 shows the distribution of longitudinal shifts for both groups of treatment plans.

## Systematic and random error

Table 1 shows the systematic error *M*, its standard deviation $\Sigma$, the random error $\sigma$, and the total error *s* of patient position in vertical (Vrt), longitudinal (Lng), lateral (Lat) direction and of the couch rotation (Rtn), with respect to the position of reference isocentre (skull/neck). Positive values refer to shifts in the inferior, cranial and left direction. In addition to the data obtained from the record & verify system, the magnitude of the 3D positional correction (*R*) was calculated, $R=\sqrt{x^2+y^2+z^2}$, where *x*, *y* and *z* are positional corrections in the longitudinal, lateral and vertical directions.

Regardless of the position of reference isocentre, systematic error *M* is in the sub-millimetre range in the longitudinal and lateral direction, while in the vertical direction it exhibits a systematic negative shift. The differences between the values obtained by different reference isocentre positions are generally an order of magnitude lower than the standard deviation of systematic error.

It can be seen that the magnitude of error depends more strongly on the direction of the shift than on the position of the reference isocentre. For both reference isocentre positions, the largest systematic error *M* is in vertical direction; in this direction one can also observe the largest standard deviation of the systematic error Σ, also for both reference isocentre positions. The standard deviation of the random error σ is largely independent of the direction. The most interesting case seems to be the systematic error in couch rotation, where the choice of reference isocentre position affects the sign of *M*.

## Positional shift distribution

Fig. 2 shows the histograms of the positional correction applied. The distributions of positional shifts in the vertical, longitudinal and lateral directions appear bell-shaped. It can also be seen that the centre of distribution for vertical positional shifts is tilted to the negative values, as already known from Table 1. An interesting feature which cannot be discerned from the table is seen in Fig. 2d. This distribution is not bell-shaped. Instead it exhibits a very sharp peak at 0°, surrounded by depleted areas on both sides of it. There are practically no rotational corrections of 0.2° or 0.3°. In both directions, there are low and broad peaks between 1° and 1.5°, and the curve falls again for larger values. It shows that radiation therapists are generally reluctant to apply rotational corrections smaller than 0.5°, rounding very small angles to 0.

Fig. 3a shows the distribution of the magnitude of positional corrections, *R*. Both distributions – with a reference isocentre on the patient head (red) and on the patient neck (blue) – have a peak in the 3-4 mm bin. Theoretically, if the positional shifts in the vertical, longitudinal and lateral directions obey a normal distribution, *R* is expected to follow a third-order χ-distribution.

## Wilcoxon rank-sum test

In each of the four degrees of freedom (three translations plus rotation), the distributions of positional shifts for the two groups of patients – those with the reference isocentre on the skull and those with the reference isocentre on the neck – were pairwise compared using the nonparametric Wilcoxon rank-sum test (Mann-Whitney *U*-test).

Wilcoxon rank-sum test shows that the distributions of positional corrections in the longitudinal direction with the reference isocentre on the skull and on the neck differ significantly ($p < 0.05$, two-tailed). In the other three comparisons of positional correction distributions, Wilcoxon rank-sum test does not show any significant difference ($p = 0.38$, $0.97$ and $0.27$ for the vertical, lateral and rotational correction, respectively).

## Correlation between positional shifts in different directions

Next, the correlations between positional corrections in different directions were examined. Fig. 4 shows all the pair-wise combinations as off-diagonal scatterplot panels in the top left triangle; e.g., the top left panel shows the correlation between the vertical (on the *x*-axis) and rotational (*y*-axis) corrections. *x*- and *y*-coordinates of an individual circle in a scatterplot correspond to the positional corrections applied in different directions at an individual treatment fraction. Red and blue circles denote the patients with reference isocentre on patient skull and patient neck, respectively. Kendall rank correlation coefficient ($\tau_B$) and the corresponding *p*-value were computed and are shown in the off-diagonal panels in the lower right triangle. Kendall's $\tau_B$ was used in place of more widely used Spearman's ρ in order to avoid the problem with ties. The absolute value of $\tau_B$ and *p* for the patients with reference isocentre on the skull and on the neck are given in red and blue, respectively. One can see that overall, positional corrections in different directions are only weakly correlated, the highest being the correlation between couch rotation and the lateral translational shift for both reference isocentre set-ups, and between longitudinal and vertical shift for the patients with the reference isocentre on patient head only.

Such low values of correlation coefficients involving couch rotation are largely influenced by the fact that with both reference isocentre set-ups, no rotational correction was performed in approximately 75% of all cases. Having the couch fixed to 0° regardless of the translational correction means that the rotational correction is completely non-correlated. If we limit this treatment only to the treatment fractions in which couch rotations were performed – this means excluding all treatment fractions with couch rotation in the [−0.1º, 0.1º] range – we find out that some correlation coefficients undergo a sizable increase. The correlation coefficients between couch rotation and lateral shift increase to 0.37 ($p < 0.01$) and 0.33 ($p < 0.01$) for the reference isocentre on the skull and on the neck, respectively. The correlation coefficients between couch rotation and longitudinal shift also increase significantly, to 0.15 ($p < 0.01$) and 0.10 ($p = 0.02$) for the reference isocentre on the skull and on the neck, respectively. On the other hand, the correlation coefficients between couch rotation and vertical shift increase only minimally, to 0.04 ($p = 0.22$) and 0.06 ($p = 0.18$), respectively, for the both reference isocentres. This can be understood, as isocentric rotation and translational shifts in the lateral and longitudinal directions are coupled transformations; you cannot perform one without affecting the other. A shift in the vertical direction, on the other hand, is independent of the rotation around the vertical axis.

Fig. 3b shows a scatterplot of the magnitude of positional corrections (*R*) and couch rotation. Again, we can see that the scatterplot is dominated by the cases in which no couch rotation was performed. The magnitude of positional correction is only weakly correlated with the couch rotation, even when excluding the cases in which no couch rotation was performed.

# Discussion

## Comparison with the published data

In an early study, portal electronic images were employed to reduce systematic set-up errors (5). With careful daily positioning using a simple head and neck fixation device, the authors obtained random patient set-up errors with an SD of 1.5 mm, whereas the SD of the systematic errors was approximately 2 mm for both directions in the sagittal plane. Using the shrinking action level (SAL) off-line set-up correction protocol (6), they were able to reduce the systematic errors to approximately 1 mm.  In another early study, the impact of the choice of thermoplastic mask on the set-up error was studied (7). The largest systematic displacement was in longitudinal direction at the shoulder level (1.6 mm) and at the head-level (1.4 mm); the rest were smaller than 0.7 mm. Standard deviations were 1.0-1.8 mm in all directions for head, neck and shoulder level.

Along with the thermoplastic mask, the choice of head support has also been given attention (8). By replacing the standard soft cushions with the ones moulded to conform to the posterior contour of the patient head, the authors were able to reduce the systematic error in the longitudinal, lateral and vertical direction from 2.2-2.3 mm to 1.2-2.0 mm. The random errors in the lateral and longitudinal direction were reduced from 1.6 mm to 1.1 mm and 1.0 mm, respectively.  Using SAL off-line set-up correction protocol, they further reduced systematic error to 0.8-1.1 mm, keeping the deviations within ±5 mm in all cases.

Table 2 presents the results of set-up error assessment from several recent studies, employing a variety of imaging techniques. For this comparison, the data for both choices of the reference isocentre were aggregated. The standard deviation of the systematic error $\Sigma$ for translational errors obtained in this study is in line with the higher values published (8-12), and the values for random error $\sigma$ fit in the middle of those published.  Less data has been published on rotational errors, but again, our values seem to be very close to the published ones (13). In order to further reduce $\Sigma$, where we seem to have some room for improvement, the experience with reducing the set-up errors in Tampere (14) may be useful.  Using simple changes in the set-up procedure like the thermoplastic mask formation and the image matching, they were able to reduce the residual systematic error by almost 50%.

Attempting to account for different mobilities of various anatomical sub-regions, a group from Amsterdam employed cone-beam CT (CBCT)  and defined 8 different regions of interests (ROIs) in addition to a large clinical ROI  (15). While $\Sigma$ and $\sigma$ for the large ROI were 1.2 mm and 1.5 mm, respectively, the values for small ROIs were generally larger. $\Sigma$ differed from 0.4 mm for C1-C3 vertebrae group to 3.8 mm for the larynx; the corresponding random error $\sigma$ ranged from 0.5 mm to 3.6 mm.

In a study comparing the set-up correction values obtained by the CBCT with the ones obtained by the 2D/2D kV match (16), the authors found a high and statistically significant correlation between the translational shifts obtained by the two methods (Spearman's rank correlation coefficient $\rho$ ranging from 0.60 to 0.72 depending on direction, $p < 0.0001$). On the other hand, the rotational corrections obtained by the two methods were found to be uncorrelated.

The online pre-treatment verification scheme with 0 mm action threshold can help reducing the systematic error in patient set-up (17). However, the increased accuracy comes at a cost of an increased workload for radiation therapists, who have to do their job in demanding time constraints. In this respect, the off-line set-up protocols are better, since image matching can be done at a pace suited to the radiation therapist. In a study designed to assess the difference between the online and off-line image matching (18), a group of radiation therapists was asked to realign the pre-treatment images with DRRs. Positional shifts obtained by the online and off-line matches were compared using the Bland-Altman formalism (19), and strong agreement between the two sets was found: with 95% probability, the difference between the online and off-line set-up shift was found to be 1.2 mm or less.

## From setup errors to safety margins

Setup errors are usually analysed with the intention of obtaining safety margin between the clinical target volume (CTV) and the planning target volume (PTV), e.g., using any of the well-known formulas (3). There are however a few issues which deserve special attention when attempting to derive safety margins from setup errors, which we will cover in the next three subsections.

### Gross errors

While there exists a general agreement that a gross error is an unacceptably large set-up error with a possibly detrimental effect on the treatment, no consensus on the threshold of what constitutes a gross error has been reached. 5 mm is often implicitly considered as a threshold for gross error (8,10,12). When considering safety margins, a value as low as 3 mm is sometimes used (20), while for the development of an automated patient safety system 10 mm was used (21).

Comparing to other studies, we observe a disproportionately large amount (28%) of gross errors (> 5 mm) in our study. We suspect the reason for this is high workload (< 15 min per patient fraction) coupled with the radiation therapists' confidence in the online imaging protocol, which allows them to correct the patient position before treatment. The few records excluded from the analysis were usually related to the omission of a step in the patient set-up procedure, e.g., failing to apply the planned shift from the reference isocentre to the isocentre. As the patient was imaged before treatment, the discrepancy was discovered before the patient was treated. However, these are an entirely different class of errors, which need to be dealt separately by employing appropriate quality assurance measures (3). As they would introduce unreasonably high values into the statistics, they were left out.

### MV and kV imaging

Early studies (5,7-10) relied on electronic portal imaging device, which shares the radiation source with the treatment system. However, with the advancement of kV imaging systems (22), the latter are being predominantly used for image-guided radiotherapy (IGRT) nowadays. In a study comparing kV and MV imaging, the authors reported a decrease in the standard deviation of systematic set-up error $\Sigma$ when using kV imaging, while maintaining the level of inter-observer variance and reducing intra-observer variance (23).

While kV imaging appears advantageous due to a lower dose it imposes onto the patient (24), one needs to keep in mind that the kV imaging system is a system separeate from the treatment system, and needs to be brought in alignment with the latter. Quality assurance recommendations (e.g., (25)) specify a 2 mm tolerance for the isocentre indicated by the IGRT system corresponding to the treatment isocentre. This needs to be kept in mind whenever the systematic and random errors are used as a basis for safety margins.

### Inter- and intrafraction shifts

Set-up errors are important information for deriving the CTV-PTV margin (3). The values for set-up errors obtained in this study are directly applicable should we decide to abolish daily imaging in favour of some average patient shift-based scheme.  If we however want to stay with the daily imaging, it is the intrafraction shifts, not the interfraction, which are more relevant for safety margins.  Assessing the intrafraction shifts involves additional patient imaging, which is a mildly invasive procedure. At present, we wanted to keep this study completely non-detrimental for the patient.

# Conclusions

To our knowledge, this is the most extensive study of head-and-neck set-up errors in terms of the number of patients enrolled, and the first which studies the impact of reference isocentre position.

We have demonstrated that bringing the reference isocentre and the treatment field isocentre closer together, longitudinally, changes the distribution of position shifts in the longitudinal direction to the degree that the Wilcoxon rank-sum test shows a statistically significant difference; the difference is nevertheless small and  in our case should not prevail over any clinical  arguments to choose either position.  All in all we can conclude that based on the results presented, neither reference isocentre set-up has shown a clear advantage over the other in terms of interfraction set-up error.

# Acknowledgements

The authors thank Dr. Valerija Žager Marciuš and Ms. Jerneja Marolt Bukovac for helpful discussion.

# Financial Support

PP acknowledges the financial support from the Slovenian Research Agency (research core funding No. P1-0389).

# Conflicts of Interest

None.

# Ethical Standards

The authors assert that all procedures contributing to this work comply with the ethical standards of the relevant national guidelines on human experimentation and with the Helsinki Declaration of 1975, as revised in 2008, and has been approved by the Protocol Review Board at the Institute of Oncology Ljubljana (ERID-KSOPKR/140, 2016-12-22) and the Ethics Committee at the Institute of Oncology Ljubljana (ERID-EK/5, 2017-01-27).

# List of figures

1. Distribution of longitudinal shifts from the reference isocentre to treatment field isocentre for patients with reference isocentre set on patient skull (red) and neck (blue). Positive and negative values denote shifts in the cranial and caudal directions, respectively.

2. Histograms of distributions of corrections applied in the vertical, longitudinal and lateral direction, and couch rotation corrections. The red line denotes the patients with reference isocentre on patient skull, blue the patients with patient origin on patient neck.

3. Histogram of the distribution of the magnitude of translational positional corrections applied (a); the correlation between the magnitude of translational correction and rotational correction (b). Red denotes the patients with reference isocentre on the skull, while blue the patients with reference isocentre on the neck.

4. Correlations between positional corrections in different directions. An individual circle in a scatterplot corresponds to the positional corrections performed in different directions at an individual treatment fraction. Red and blue circles (blue plotted on top of red) denote the patients with reference isocentre on patient skull and patient neck, respectively. The absolute value of Kendall rank correlation coefficient ($\tau_B$) and the corresponding *p*-value for the patients with reference isocentre on the skull and on the neck are written in red and blue, respectively.

## List of tables

1. Systematic and random errors in patient position with respect to the position of reference isocentre. $M$, $\Sigma$, $\sigma$, and $s$ denote the systematic error, its standard deviation, random error, and total error, respectively. Vrt, Lng, Lat and Rtn denote the vertical, longitudinal, and lateral direction, and couch rotation. All translational positional shifts are expressed in mm, and the rotation in degrees.

|     | $M_{skull}$ | $M_{neck}$ | $\Sigma_{skull}$ | $\Sigma_{neck}$ | $\sigma_{skull}$ | $\sigma_{neck}$ | $s_{skull}$ | $s_{neck}$ |
|-----|------|------|------|------|------|------|------|------|
| Vrt | -1.4 | -1.4 | 2.5  | 2.5  | 1.8  | 1.9  | 2.9  | 3.0  |
| Lng | 0.6  | 0.9  | 2.1  | 1.8  | 1.5  | 1.6  | 2.5  | 2.3  |
| Lat | 0.6  | 0.7  | 2.9  | 1.7  | 1.6  | 1.5  | 2.4  | 2.2  |
| Rtn | 0.02 | -0.05| 0.43 | 0.49 | 0.62 | 0.60 | 0.74 | 0.76 |
| $R$ | 4.4  | 4.4  | 1.5  | 1.3  | 1.7  | 1.6  | 2.1  | 2.0  |

2. Comparison of the values of systematic ($\Sigma$) and random error ($\sigma$) in the vertical, longitudinal and lateral direction, and couch rotation, in a few recent studies analysing set-up errors in treating head-and-neck cancer.

| Study | $\Sigma$ (mm) | | | | $\sigma$ (mm) | | | | Imaging technique |
|---|---|---|---|---|---|---|---|---|---|
|   | Vrt | Lng | Lat | Rtn | Vrt | Lng | Lat | Rtn |   |
| Suzuki (9)  | 0.7-1.1 | 1.1-1.4 | 1.0-1.2 |      | 0.8-1.3 | 1.0-1.4 | 0.8-1.6 |      |         |
| Gupta (10)  | 0.96    | 1.20    | 0.96    |      | 1.94    | 2.48    | 1.97    |      | MV 2D   |
| Štrbac (11) | 1.42    | 1.52    | 1.93    |      | 1.77    | 1.83    | 1.83    |      | kV 2D   |
| Zumsteg (12)| 1.8     | 2.8     | 1.9     |      | 2.6     | 2.2     | 2.0     |      | MV CBCT |
| Oh (13)     | 1.0     | 0.8     | 1.1     | 0.8° | 1.2     | 1.1     | 1.3     | 0.6° | kV CBCT |
| This study  | 2.5     | 2.0     | 1.8     | 0.5° | 1.8     | 1.5     | 1.6     | 0.6° | kV 2D   |

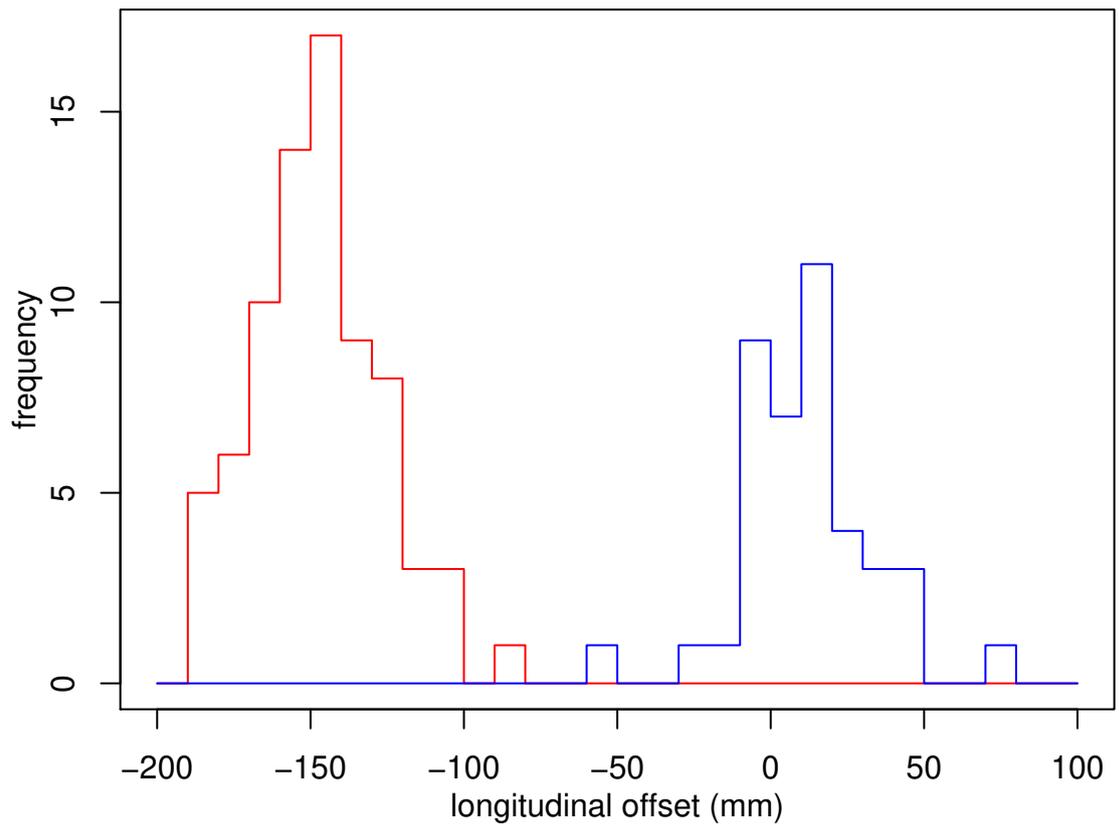

Figure 1

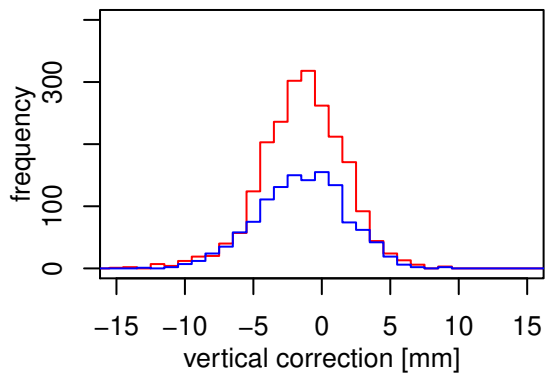
(a)

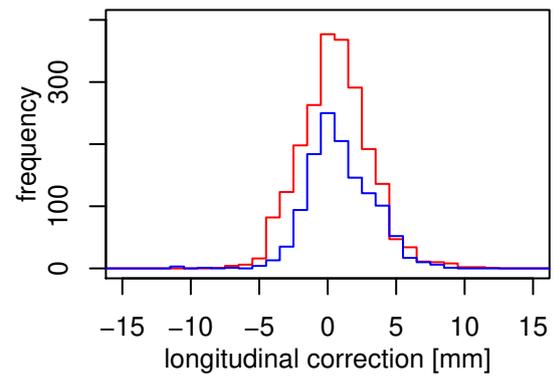
(b)

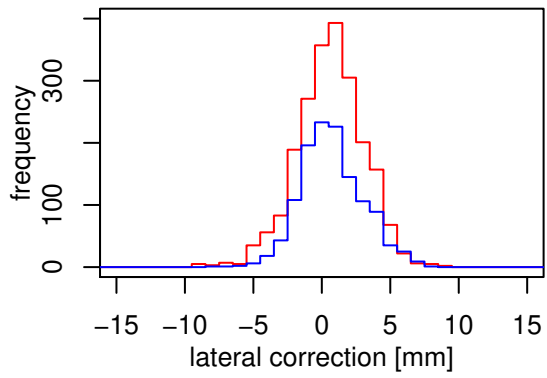
(c)

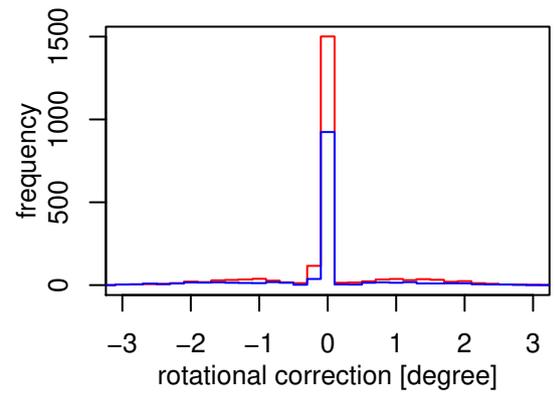
(d)

Figure 2

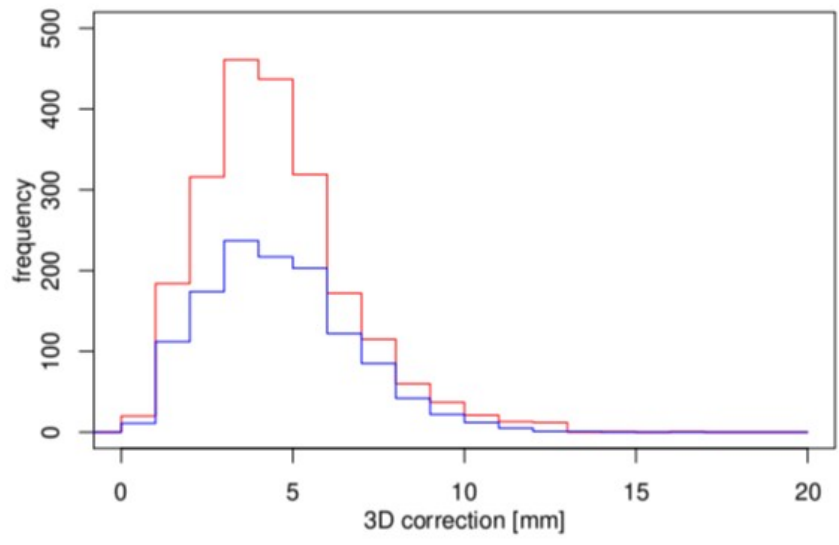

(a)

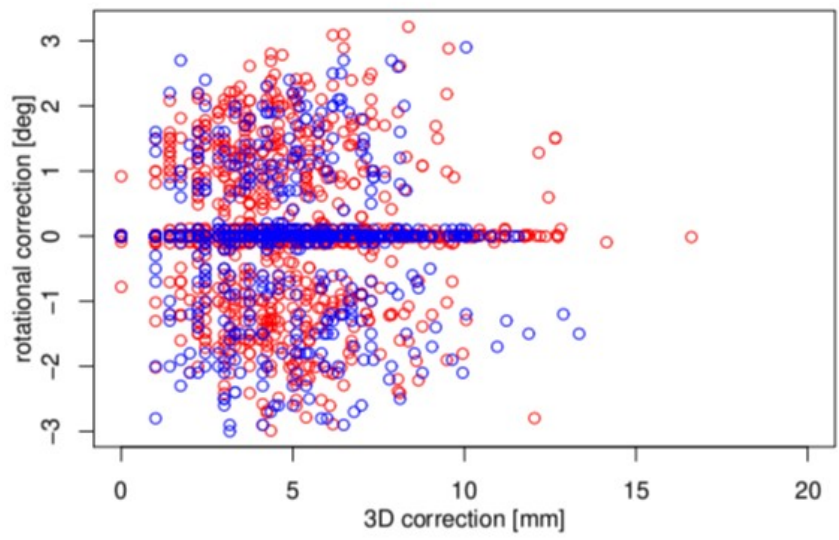

(b)

Figure 3

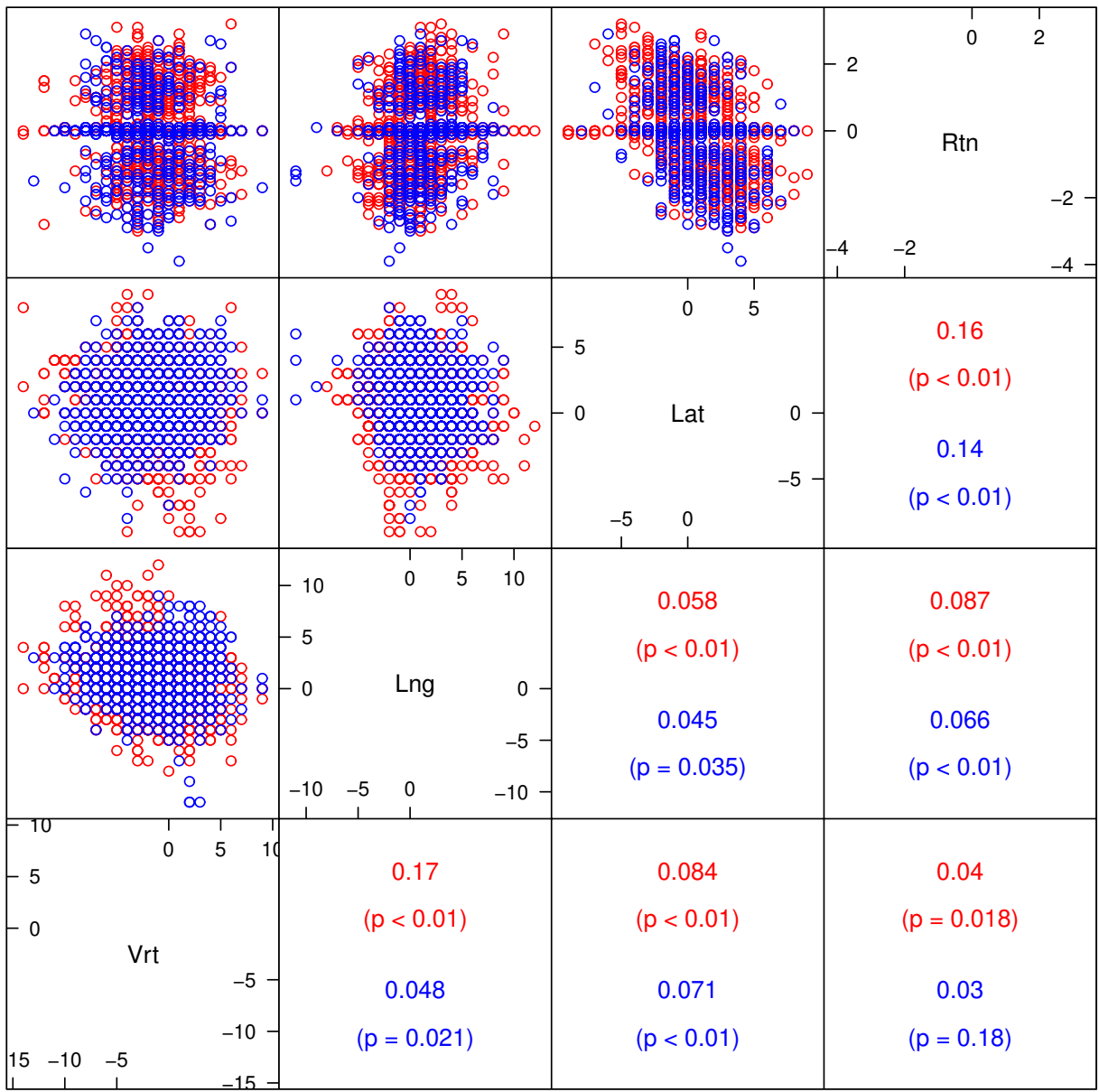

Figure 4